\documentclass[floats,aps,amssymb,nofootinbib,prd]{revtex4}
\usepackage{amssymb}
\usepackage{graphics}
\usepackage{amsmath}
\usepackage{amsfonts}
\usepackage{bm}
\usepackage[]{latexsym}
\usepackage{float}
\usepackage{graphicx}
\usepackage{color}

\newcommand{\be}{\begin{equation}}\newcommand{\ee}{\end{equation}}
\newcommand{\bea}{\begin{eqnarray}}\newcommand{\eea}{\end{eqnarray}}
\newcommand{\brr}{\begin{array}}\newcommand{\err}{\end{array}}
\newcommand{\bit}{\begin{itemize}}\newcommand{\eit}{\end{itemize}}
\newcommand{\ben}{\begin{enumerate}}\newcommand{\een}{\end{enumerate}}

\newcommand{\ba}{\begin{array}}
\newcommand{\ea}{\end{array}}

\def\lf{\left}

\def\non{\nonumber}

\def\ri{\right}
\def\al{\alpha}
\def\de{\delta}

\def\si{\sigma}\def\Si{\Sigma}

\def\1{{_{1}}}\def\2{{_{2}}}

\def\noHe0{:\;\!\!\;\!\!:H_e(0):\;\!\!\;\!\!:}
\def\noHm0{:\;\!\!\;\!\!:H_\mu(0):\;\!\!\;\!\!:}

\def\lf{\left}

\def\non{\nonumber}

\def\ri{\right}

\def\al{\alpha}
\def\de{\delta}

\def\si{\sigma}\def\Si{\Sigma}

\def\1{{_{1}}}\def\2{{_{2}}}

\begin{document}
\title{Gravitationally modulated quantum correlations: Discriminating classical and quantum models of ultra-compact objects with Bell nonlocality}

\author{Luciano Petruzziello\footnote{lupetruzziello@unisa.it}$^{\hspace{0.3mm}1,2,3}$ and Fabrizio Illuminati\footnote{filluminati@unisa.it}$^{\hspace{0.3mm}1,2}$} \affiliation
{$^1$Dipartimento di Ingegneria Industriale, Universit\`a degli Studi di Salerno, Via Giovanni Paolo II, 132 I-84084 Fisciano (SA), Italy. \\
$^2$INFN, Sezione di Napoli, Gruppo collegato di Salerno, Fisciano (SA), Italy. \\
$^3$Institut f\"ur Theoretische Physik, Albert-Einstein-Allee 11, Universit\"at Ulm, 89069 Ulm, Germany.}

\date{\today}
\def\be{\begin{equation}}
\def\ee{\end{equation}}
\def\al{\alpha}
\def\bea{\begin{eqnarray}}
\def\eea{\end{eqnarray}}

\date{April 21, 2023}

\begin{abstract}
We investigate the relation between quantum nonlocality and gravity at the astrophysical scale, both in the classical and quantum regimes. Considering a gedanken experiment involving particle pairs orbiting in the strong gravitational field of ultra-compact objects, we find that the violation of Bell inequality acquires an angular modulation factor that strongly depends on the nature of the gravitational source. We show how such gravitationally-induced modulation of quantum nonlocality readily discriminates between black holes (both classical and inclusive of quantum corrections) and string fuzzballs, i.e., the true quantum description of ultra-compact objects according to string theory. These findings promote Bell nonlocality as a potentially key tool in putting quantum gravity to the test.
\end{abstract}

\vskip -1.0 truecm 

\maketitle

\section{Introduction}

The development of a consistent and predictive theory of quantum gravity is one of the main unresolved conundrums in contemporary physics~\cite{loll}. Relentless efforts in the attempt to reconcile quantum mechanics and general relativity have produced a number of promising candidate models, including asymptotic safety~\cite{ass}, causal dynamical triangulations~\cite{causal}, non-commutative geometry~\cite{ncg}, loop quantum gravity~\cite{lqg}, doubly special relativity~\cite{dsr} string theory~\cite{st} {and the more recent proposal of ``gravitizing'' quantum mechanics \cite{gravquant}.}
All the aforementioned theoretical schemes have their own characteristics and predictions which make them profoundly different among each other. Despite that, it is still possible to recognize similar aspects which are thus likely to be part of a general treatment of quantum gravity {(see for instance Refs. \cite{rec1,rec2} and therein for a review on this topic)}. {Prominent examples of features foreseen by many of the above models are the emergence of an intrinsic non-local behavior in the theoretical description of quantum gravity (\emph{i.e.}, see Refs. \cite{nl1,nl2}) and} the existence of a minimal length at the Planck scale with the ensuing modifications of the canonical commutation relations of quantum mechanics and the associated Heisenberg uncertainty principle~\cite{gup1,gup2}.

{Concerning the notion of a minimal spatial resolution, this} can be deduced also from gedanken experiments involving large~\cite{large} and micro~\cite{micro} black holes, in proximity of which quantum gravitational effects are expected to become dominant. As a matter of fact, the strong gravity regime near a black hole\footnote{{For the purpose of the present analysis, henceforth we assume to work with the strong gravity regime comprising the region of spacetime in proximity of the horizon.}} prevents the use of any known approximation in the study of quantum systems. {To quote a relevant example along this direction, it is worth observing that, although the extension of quantum field theory to curved backgrounds has provided successful fundamental predictions (such as the Hawking radiation), these findings are still plagued by unphysical divergences when considered beyond their limits of applicability. However, this fact does not undermine the validity of the aforementioned results, but it rather points towards the quest for a unified description of quantum and gravitational phenomena.} Achievements in addressing these difficulties would yield major progress towards a viable theory of quantum gravity able to settle open issues such as the information paradox and the singularity problem that arise in the context of classical and semi-classical approaches to gravitational phenomena. 

An interesting resolution for both of the above issues in the framework of superstring theory is represented by the fuzzball proposal~\cite{fuzz1,fuzz2}, according to which the supposed black hole is in fact conceived as a massive object made of a very large number of microscopic strings which, by definition, feature a minimal length extension qualitatively of the order of the Planck scale. Even though the original arguments leading to the fuzzball solution were purely theoretical, it has been recently pointed out that concrete realizations of fuzzballs lead to a phenomenology that might be accessible, for instance via the observational investigation of gravitational waves~\cite{bianchi}.

{Nevertheless, the fuzzball proposal is not the only self-consistent and robust alternative concerning the generalization of black hole physics that incorporates quantum gravitational effects. Indeed, also loop quantum gravity predicts the settlement of the problems discussed above by relying on the underlying spacetime discretization \cite{lqgbh}, whilst for the asymptotic safety paradigm the solution is to be found in the quantum scale invariance \cite{asbh}. On a final note, it is worth stressing that another important class of black holes can be derived in the context of higher-derivative theories and non-local gravity \cite{hdgbh}. These models cure the issues stemming from merging quantum field theory and gravitation (\emph{i.e.}, non-renormalizability, non-unitarity, etc.) by adding higher-derivative terms in the Lagrangian of the gravitational interaction. Interestingly, these contributions are able to remove the unwanted features that affect the canonical quantization of Einstein's general relativity and may give rise to potentially detectable effects in a significant number of physical phenomena.}

In a parallel development, the community active in quantum information science, atomic physics and quantum optics has picked up in recent years on the original ideas by Bronstein and Feynman \cite{Bronstein1,Bronstein2,Feynman}, suggesting to test the hypothetical quantum nature of gravity in the laboratory by measuring witnesses of the bipartite entanglement between two test masses induced by a quantized gravitational field, i.e. a quantum gravitational mediator \cite{Bose,Marletto}.  

Motivated by the above considerations, in the present work we address the broader question of gravitationally-induced modifications of nonlocal quantum correlations. Given that a classical gravitational mediator cannot induce any form of quantum nonlocality, be it, in ascending hierarchical order, entanglement, steering, or Bell nonlocality, we investigate whether classical and quantum gravity can have different effects on {\em{already}} existing quantum correlations, previously established by other physical interactions on pairs of test masses. To this end, we study a gedanken experiment which revolves around the dynamics of the Bell nonlocality of particle pairs in the gravitational field generated by ultra-compact objects of diverse nature, such as black holes and fuzzballs, in order to assess whether and how different gravitational sources affect the dynamical evolution of quantum nonlocality. 

Historically, establishing a relation between cosmological objects and quantum entanglement was the central result of a celebrated paper by Maldacena and Susskind~\cite{epepr}, where it was conjectured that the entanglement shared by two particles can be interpreted as a non-traversable wormhole; such a correspondence may be viewed as a precondition for the unification of quantum and gravitational effects. 
{In addition to the aforesaid achievement, the interplay between gravity and entanglement can be identified in a significant number of relevant frameworks. For instance, it is worth recalling that, by means of entanglement entropy, it is possible to deduce another theoretical evidence of black hole thermodynamics related to the area law \cite{arealaw}. Along the same direction, by means of thermodynamical arguments one can also show that Einstein's field equations of general relativity must necessarily be fulfilled if entanglement equilibrium is established \cite{thermogr}.}

Here, instead, by relying on Einstein-Podolsky-Rosen (EPR) nonlocal correlations~\cite{epr} shared by particle pairs orbiting around ultra-compact objects, we investigate what insights Bell nonlocality, rather than entanglement, can provide on the nature and properties of gravitational structures. In this respect, it is important to recall once more that nonlocality and entanglement are distinct concepts that stand in a hierarchical relation: whilst a violation of Bell inequality always implies entanglement, the opposite implication does not necessarily hold, a notorious counterexample being that of the Werner mixed two-qubit states~\cite{werner}, which can be entangled without violating Bell inequality.

Proceeding to evaluate explicitly the amount of Bell nonlocality in an extreme astrophysical scenario, we resort to the physically transparent Clauser-Horne-Shimony-Holt (CHSH) form of Bell inequality~\cite{chsh,bell,bellreview} for massive spin-$1/2$ particle pairs, and we find that gravity in the strong-field regime affects significantly the quantum nonlocality shared by the test particles. Indeed, the overall degree of violation of the CHSH inequality is modulated by an angular factor that strictly depends on the nature of the ultra-compact object under consideration. 

This result, which is completely general and may be adapted also to different frameworks, is elucidated by focusing on three relevant cases: the classical Schwarzschild black hole, the Schwarzschild black hole within a quantum-corrected treatment at leading (perturbative) order and the string fuzzball solution. We find that the gravitational modulation of bipartite Bell nonlocality discriminates unambiguously between all of them.
In order to proceed in our investigation of the thought experiment, we make use of some recently introduced techniques that allow to evaluate EPR correlations in different gravitational scenarios~\cite{ueda,adami,our}; as a side result of the main analysis, we generalize such techniques and extend their range of validity to include any static and spherically symmetric spacetime whose metric tensor is expressed in isotropic coordinates.

The paper is organized as follows: in Sec. II we introduce the necessary mathematical tools based on the concept of Wigner rotation in curved spacetime, as it is needed in the analysis of the EPR correlations shared by two spin-$1/2$ particles in the gravitational field of an ultra-compact object. Section III is devoted to the explicit computation of the Wigner rotation in various regimes; this result is then applied to the evaluation of the EPR correlations in Sec. IV. In Sec. V we discuss and compare three relevant instances of static and spherically symmetric ultra-compact gravitational objects, i.e. the string  fuzzball, the classical Schwarzschild black hole and the quantum-corrected Schwarzschild black hole, and we show how for each of them the orbiting particle pairs feature a different degree of Bell nonlocality. Finally, in Sec. VI we comment on our results and perspectives on future research.

Throughout the manuscript, we adopt Planck units ($c=\hbar=G=1$) and the mostly positive signature for the metric, i.e., $\eta_{ab}=\mathrm{diag}(-,+,+,+)$.

\section{Wigner rotation in curved spacetime}

For a consistent treatment of spin-$1/2$ particles in curved spacetime it is necessary to make use of the tetrad (or vierbein) formalism; for a comprehensive introduction on this subject, the interested reader can consult Ref.~\cite{gravitation}. A tetrad field $e^\mu_a$ evaluated at a spacetime point $x$ is completely characterized by the relation
\be\label{tetrad}
g_{\mu\nu}(x)e^\mu_a(x)e^\nu_b(x)=\eta_{ab} \, ,
\ee
where the summation over repeated indexes is understood, $g_{\mu\nu}(x)$ is the metric tensor defined on the Riemannian manifold and $\eta_{ab}$ is the Minkowski metric acting on the flat plane tangent to the manifold in the point $x$. Henceforth, to discriminate between the indexes of the manifold and of the tangent bundle, we employ Greek letters for the former and Latin letters for the latter.

The expression~\eqref{tetrad} is essential to analyze spin-$1/2$ particle states in curved backgrounds, since they are defined as the states that belong to the spin-$1/2$ representation of the local Lorentz transformation (LLT) group, while general relativity is based upon invariance under diffeomorphisms. Precisely in order to build a bridge between these two notions, one can introduce tetrads, as they allow to ``project'' diffeomorphism-covariant tensors of the differentiable manifold onto local Lorentz-covariant quantities defined on a flat tangent plane. 

By virtue of this procedure, a generic spin state with four-momentum $k^\mu=mu^\mu$ (where $u^\mu u_\mu=-1$) at the spacetime point $x$ can be unambiguously labeled with $|k^a,\si;x \rangle$, where $k^a=e^a_\mu k^\mu$ and $\si=\uparrow,\downarrow$ is the third component of the spin. Naturally, the field $e^a_\mu$ is the inverse of the one appearing in Eq.~\eqref{tetrad}; consequently, the following identities hold:
\be\label{inverse}
e^\mu_a e^a_\nu=\delta^\mu_\nu\,, \qquad e^\mu_a e^b_\mu=\delta^b_a\,.
\ee
If we now want to describe the dynamical evolution of a spin-$1/2$ particle moving in curved spacetime, we have to account for different flat tangent spaces, each of which is associated to a given point of the particle's trajectory. As a first step, we consider what happens after an infinitesimal interval of proper time $d\tau$, after which the particle is located at the new point $x'^\mu=x^\mu+u^\mu d\tau$. Accordingly, the shift in momentum is given by
\be\label{mom}
k^a(x')=k^a(x)+\delta k^a(x)\,,
\ee
where the variation is made of two distinct contributions, namely\footnote{When there is no need for disambiguation, the dependence on the spacetime position will be omitted.}:
\be\label{mom2}
\delta k^a=\delta k^\mu e^a_\mu+k^\mu\delta e^a_\mu \, .
\ee
By defining the four-acceleration $a^\mu=u^\nu\nabla_\nu u^\mu$ originated by an external force and recalling that $k^\mu k_\mu=-m^2$ as well as $k^\mu a_\mu=0$, it is straightforward to observe that the first variation of Eq.~\eqref{mom2} becomes 
\be\label{mom3}
\delta k^\mu=ma^\mu d\tau=-\frac{1}{m}\lf(a^\mu k_\nu-k^\mu a_\nu\ri)k^\nu d\tau \, .
\ee
On the other hand, the second factor of Eq.~\eqref{mom2} can be rewritten by introducing the expression for the connection one-form~\cite{gravitation}, that is, $\omega^a_{\mu b}=e^a_\nu\nabla_\mu e^\nu_b$, and hence 
\be\label{mom4}
\delta e^a_\mu=-u^\nu\omega^a_{\nu b}e^b_\mu d\tau=\xi^a_b e^b_\mu d\tau\,.
\ee 
In so doing, exploiting Eqs.~\eqref{mom3} and~\eqref{mom4} to rewrite Eq.~\eqref{mom2}, one can identify an infinitesimal local Lorentz transformation occurring for the quantity $k^a$. As a matter of fact
\be\label{llt}
\delta k^a=\lambda^a_b k^bd\tau \, ,
\ee
where 
\be\label{llt2}
\lambda^a_b=-\frac{1}{m}\lf(a^ak_b-k^aa_b\ri)+\xi^a_b
\ee
is an infinitesimal LLT. This means that the momentum of the particle as viewed by a local reference frame (i.e., the one belonging to the tangent space) undergoes the transformation
\be\label{llt3}
k^a(x')=\Lambda^a_b(x)k^b(x)\,, \qquad \Lambda^a_b=\delta^a_b+\lambda^a_b \, ,
\ee
which is precisely a LLT. Consequently, the evolution of a spin-$1/2$ state must be described in terms of a representation of the spin-$1/2$ local Lorentz group. Bearing this in mind, we recall that, in the context of flat spacetime, under the action of a given Lorentz transformation $\Lambda^a_b$, the spin-$1/2$ one-particle state $|k^a,\si\rangle$ transforms as follows~\cite{tung,weinberg}:
\be\label{trasf}
U(\Lambda)|k^a,\si\rangle=\sum_{\si'}D_{\si'\si}^{(1/2)}\lf(W\lf(\Lambda,k\ri)\ri)|\Lambda k^a,\si'\rangle \, ,
\ee
with $D_{\si'\si}^{(1/2)}\lf(W\lf(\Lambda,k\ri)\ri)$ being a $2\times2$ unitary matrix that allows for the the Wigner rotation $W^a_b(\Lambda,k)$ of the spin. The Wigner rotation~\cite{wigner} can be written as 
\be\label{wig}
W^a_b(\Lambda,k) =\lf[L^{-1}(\Lambda k) \Lambda L(k)\ri]^a_b \, ,
\ee
where $L^a_b$ is the Lorentz boost
\be\label{lb}
L^0_0=\Xi\,, \quad L^i_0=L^0_i=\frac{k^i}{m}\,, \quad L^i_j=\de_{ij}+\lf(\Xi-1\ri)\frac{k^ik^j}{|\vec{k}|^2} \, ,
\ee
with $\Xi=\sqrt{|\vec{k}|^2+m^2}$ and the indexes $i,j=1,2,3$.

When generalizing to include the case of a curved spacetime, we have to resort to local Wigner rotations stemming from the LLTs described in Eq.~\eqref{llt3}. Accordingly, Eq.~\eqref{wig} becomes
\be\label{wig2}
U(\Lambda(x))|k^a,\si;x\rangle=\sum_{\si'}D_{\si'\si}^{(1/2)}\lf(W(x)\ri)|\Lambda k^a,\si';x\rangle \, .
\ee
Notice that a similar scenario holds true not only for spinors, but for Dirac bispinors as well; for recent applications of the latter, see Refs.~\cite{bispinors} and references therein.

The form of the infinitesimal local Wigner rotation can be extracted from Eq.~\eqref{llt3}; indeed, one can verify that
\be\label{wig3}
W^a_b=\delta^a_b+\vartheta^a_b d\tau\,, \qquad \vartheta^i_j=\lambda^i_j+\frac{\lambda^i_0k_j-\lambda_{j0}k^i}{k^0+m} \, ,
\ee
where $i,j=1,2,3$, as they are the only non-vanishing terms of $\vartheta^a_b$.


\section{Wigner rotation for a generic metric in isotropic coordinates}

In the following, we compute the Wigner rotation angle for a general class of static and spherically symmetric spacetime solutions, thus going beyond the standard Schwarzschild case treated in Ref.~\cite{ueda} and the weak-field limit considered in Ref.~\cite{our}. To this aim, we make use of a generic line element that can be cast in isotropic spherical coordinates as follows:
\be\label{metric}
ds^2=-f(r)dt^2+g(r)\lf[dr^2+r^2d\theta^2+r^2\sin^2\theta d\varphi^2\ri] \, .
\ee
We note in passing that, by setting $f(r)=(1-M/2r)^2/(1+M/2r)^2$ and $g(r)=(1+M/2r)^4$, one recovers the results of Ref.~\cite{ueda} within a different coordinate system, while if $f(r)=1+2\phi(r)$ and $g(r)=1-2\psi(r)$, with $\phi(r)$ and $\psi(r)$ being weak gravitational potentials arising in extended theories of gravity, one recovers the findings of Ref.~\cite{our}. 

As the metric tensor is diagonal, we can compute the tetrads rather straightforwardly
\be\label{vier}
e^t_0=\frac{1}{\sqrt{f}}\,, \quad e^r_1=\frac{1}{\sqrt{g}}\,, \quad e^\theta_2=\frac{1}{r\sqrt{g}}\,, \quad e^\varphi_3=\frac{1}{r\sin\theta\sqrt{g}} \, .
\ee
From the above equation, it is possible to deduce the non-vanishing components of the connection one-form
\bea\non
\omega^0_{t1}=\omega^1_{t0}=\frac{f'}{2\sqrt{gf}}\,, \qquad \omega^2_{\theta1}=-\omega^1_{\theta2}=1+\frac{rg'}{2g}\,, \qquad \omega^3_{\theta3}=\cot\theta\,,\\[2mm]\label{oneform}
\omega^3_{\varphi1}=-\omega^1_{\varphi3}=\lf(1+\frac{rg'}{2g}\ri)\sin\theta\,, \qquad \omega^3_{\varphi2}=-\omega^2_{\varphi3}=\cos\theta\,, \qquad
\eea
where the prime denotes derivation with respect to the coordinate $r$.

\begin{figure}[h]
\centering
\includegraphics[width=11.5cm]{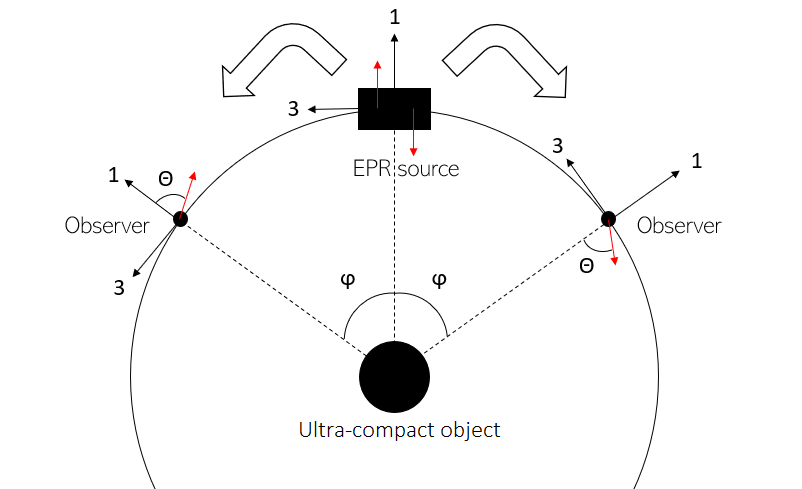}
\caption{A pair of spin-$1/2$ particles initially sharing a perfect EPR correlation is produced at $\varphi=0$. Particles travel along a circular orbit around the ultra-compact object in opposite directions. At the end of each propagation, the spins are rotated due to the presence of a non-trivial background spacetime.}
  \label{figure}
\end{figure}

Without loss of generality, we can investigate the circular motion\footnote{{In principle, one could analyze geodesic orbits and/or other kind of closed trajectories in proximity of the ultra-compact object. However, the circular motion has been selected to streamline computations, but such a peculiar choice does not spoil the qualitative behavior of the final result.}} of the entangled particles around the ultra-compact object by assuming that the dynamics takes place on the equatorial plane $\theta=\pi/2$. Additionally, we let the EPR source be located at $\varphi=0$ and the two observers performing the local spin measurements at $\pm\varphi$. A sketch of the physical setup is shown in Fig.~\ref{figure}.

Due to the above requirements, the expression of the four-velocity is simplified, as
\be\label{vel}
u^\mu(x)=\lf(\frac{\cosh\zeta}{\sqrt{f}},0,0,\frac{\sinh\zeta}{r\sqrt{g}}\ri)\,,
\ee
where $\zeta$ denotes the rapidity in the local reference frame.

We recall that the motion under investigation is not a geodesic one; therefore, there must be an external force acting on the system that perfectly compensates the presence of gravity {and prevents the emergence of geodesic instabilities in proximity of the horizon}. Such a force produces the non-vanishing acceleration
\be\label{acc}
a^\mu=\lf(0,\frac{1}{g}\lf[\frac{f'}{2f}\cosh^2\zeta-\lf(\frac{1}{r}+\frac{g'}{2g}\ri)\sinh^2\zeta\ri],0,0\ri)\,.
\ee
We can now compute the Wigner angle introduced in Eq.~\eqref{wig3}. First, by means of simple algebraic manipulations one determines the quantities $\xi^a_b$ in Eq.~\eqref{mom4}, finding the following non-vanishing components:
\be\label{xi}
\xi^1_0=\xi^0_1=-\frac{f'}{2f\sqrt{g}}\cosh\zeta\,, \qquad \xi^1_3=-\xi^3_1=\lf(\frac{2g+rg'}{2rg\sqrt{g}}\ri)\sinh\zeta \, .
\ee
By virtue of the above expressions, we can write the infinitesimal LLT~\eqref{llt2} explicitly. Indeed, recalling that $k^a=mu^a$ and $u^a=e^a_\mu u^\mu$, then $u^a=\lf(\cosh\zeta,0,0,\sinh\zeta\ri)$. In a similar fashion, $a^a=e^a_\mu a^\mu$, and thus the only non-vanishing component of $a^a$ is $a^1=e^1_r a^r=\sqrt{g} a^r$. Therefore,
\be\label{lam}
\lambda^1_0=\lambda^0_1=-\frac{1}{r\sqrt{g}}\lf[1+\frac{rg'}{2g}-\frac{rf'}{2f}\ri]\cosh\zeta\sinh^2\zeta\,, \qquad \lambda^1_3=-\lambda^3_1=\frac{1}{r\sqrt{g}}\lf[1+\frac{rg'}{2g}-\frac{rf'}{2f}\ri]\cosh^2\zeta\sinh\zeta\,.
\ee
After the evaluation of the infinitesimal LLT, the only quantity left to compute is the infinitesimal Wigner rotation~\eqref{wig3}. Because of the choice of the physical setup summarized in Fig.~\ref{figure}, the only terms of $\vartheta^a_b$ different from zero are
\be\label{theta}
\vartheta^3_1=-\vartheta^1_3=-\frac{1}{r\sqrt{g}}\lf[1+\frac{rg'}{2g}-\frac{rf'}{2f}\ri]\cosh\zeta\sinh\zeta\,.
\ee
Next, we consider the finite transformation as a Dyson series of infinitesimal ones~\cite{ueda}, whose formal sum reads 
\be\label{fwig}
W^3_1=T \exp\lf[\int^{\tau_f}_{\tau_i}\vartheta^3_1\,d\tau'\ri]=\exp\lf[\vartheta^3_1\lf(\tau_f-\tau_i\ri)\ri]\,,
\ee
where $T$ denotes the time ordering operator. 


\section{Gravitationally-induced modulation of Bell nonlocality}

According to the setting of our gedanken experiment, the EPR source emits a pair of particles, $A$ and $B$, moving away from the source in opposite directions with constant four-momenta $k^a_\pm=(m\cosh\zeta,0,0,\pm m\sinh\zeta)$ after having been prepared in the maximally entangled spin singlet
\be\label{instate}
\vert \psi \rangle = \frac{1}{\sqrt{2}}\lf(|k^a_+,\uparrow;\varphi=0\rangle_A |k^a_-,\downarrow;\varphi=0\rangle_B-|k^a_+,\downarrow;\varphi=0\rangle_A |k^a_-,\uparrow;\varphi=0\rangle_B \ri) \, .
\ee
The CHSH inequality~\cite{chsh} and the associated CHSH measurements are a powerful toolbox to access and test the degree of quantum nonlocality in the correlations between two dichotomous variables; for the problem at hand, such variables are the spins of the entangled particles. 

As a key ingredient, we need two sets of measurements $\{ \hat{A}_1, \hat{A}_2 \}$ and $\{ \hat{B}_1, \hat{B}_2 \}$ performed on parties $A$ and $B$, respectively, with the aim of detecting the orientation of the third component of the spin. If correlations of the spins in a given shared state are local in the sense of Bell theorem~\cite{bell,bellreview}, then the inequality~\cite{chsh,bellreview}
\begin{equation}
\label{CHSHQuant}
\mathcal{S}[\vert \Psi \rangle] = \vert \langle \hat{A}_1 \hat{B}_1 \rangle + \langle \hat{A}_1 \hat{B}_2 \rangle +\langle \hat{A}_2 \hat{B}_1 \rangle -\langle \hat{A}_2 \hat{B}_2 \rangle \vert \le 2\,,
\end{equation}
holds, where $\langle \hat{A}_i \hat{B}_j \rangle = \langle \Psi  \vert \hat{A}_i \hat{B}_j \vert \Psi  \rangle $. If Eq.~\eqref{CHSHQuant} is violated, spin correlations are nonlocal and local hidden-variable theories are falsified. 

Together with the state described in Eq.~\eqref{instate}, the employment of the observables
\begin{equation}
\label{Observables}
\hat{A}_1 =\hat{\Si}_x^{(A)}\,, \quad \hat{A}_2=\hat{\Si}_y^{(A)}\,, \quad
\hat{B}_1 = -\frac{\hat{\Si}_x^{(B)} + \hat{\Si}_y^{(B)}}{\sqrt{2}}\,, \quad \hat{B}_2 =-\frac{\hat{\Si}_x^{(B)} - \hat{\Si}_y^{(B)}}{\sqrt{2}}\,,
\end{equation}
allows to reach the maximum violation of the inequality allowed by quantum mechanics, namely $\mathcal{S}[\vert \psi \rangle] = 2 \sqrt{2}$, also known as the Tsirelson bound \cite{tsir}. 

Now, the maximally entangled initial state evolves in a curved spacetime, and because of the Wigner rotation the spins of the entangled particles undergo a precession motion that prevents the perfect EPR correlation of the initial state from being preserved. Clearly, 
we expect that whether and how much the propagation of the particles along a closed path will change the orientation of the spins and, in turn, the violation of the CHSH inequality, should depend on the nature of the gravitational object around which the particles are orbiting.

Now, assume that, after a finite proper time $\tau_f-\tau_i={r\sqrt{g}\,\varphi}/\sinh\zeta$, 
particles $A$ and $B$ have reached their respective detection points; in this proper time interval, the Wigner transformation can be viewed as a rotation about the 2-axis~\cite{ueda,adami,our}
\be\label{wignermatrix}
\mathbb{W}\lf(\pm\varphi\ri)=\begin{pmatrix} 1 & 0 & 0 & 0 \\ 0 & \cos\Theta & 0 & \pm\sin\Theta \\ 0 & 0 & 1 & 0 \\ 0 & \mp\sin\Theta & 0 & \cos\Theta \end{pmatrix}\,.
\ee
where $\Theta$ can be derived from Eq.~\eqref{fwig}
\be\label{btheta}
\Theta=\frac{r\sqrt{g}\,\varphi}{\sinh\zeta}\vartheta^1_3=\varphi\cosh\zeta\lf[1+\frac{rg'}{2g}-\frac{rf'}{2f}\ri]\,.
\ee
The physical meaning of the rotation angle $\Theta$ and the spin precession can be readily visualized from Fig.~\ref{figure}.

Having the explicit expression for the Wigner rotation, the transformation acting on the spin states can be computed as shown in Eq.~\eqref{wig2}. Specifically, one can verify that~\cite{ueda,adami,our} 
\be\label{spinhalf}
D_{\si'\si}^{(1/2)}=e^{\mp i\frac{\si_y}{2}\Theta}\,,
\ee
with $\si_y$ being the Pauli matrix with imaginary entries. 

Crucially, we see that, as the particles progress travelling along the orbit that circles the ultra-compact object, the initial spin-singlet state gets embroiled in a linear superposition with the spin-triplet states, which implies that measurements of the spin along the same direction are no longer perfectly correlated in the local reference frame for $\pm\varphi$~\cite{ueda,adami,our}. 

In order to preserve perfect correlation in the local reference frame, it is sufficient to rotate the bases by $\mp\varphi$ while keeping the 2-axis fixed in the point that is denoted by $\pm\varphi$. In so doing, we obtain
\be \label{newstate1}
|k^a_\pm,\uparrow;\pm\varphi\rangle'=
\cos\frac{\varphi}{2}|k^a_\pm,\uparrow;\pm\varphi\rangle\pm\sin\frac{\varphi}{2}
|k^a_\pm,\downarrow;\pm\varphi\rangle \, ,
\ee
and
\be \label{newstate2}
|k^a_\pm,\downarrow;\pm\varphi\rangle'=
\mp\sin\frac{\varphi}{2}|k^a_\pm,\uparrow;\pm\varphi\rangle+\cos\frac{\varphi}{2}
|k^a_\pm,\downarrow;\pm\varphi\rangle\, ,
\ee
so that the evolved state reads
\be\label{finstate}
\vert \psi^\prime \rangle = \frac{1}{\sqrt{2}}\Bigl[\cos\Delta\Bigl(|k^a_+,\uparrow;\varphi\rangle'|k^a_-,\downarrow;-\varphi\rangle'-|k^a_+,\downarrow;\varphi\rangle'|k^a_-,\uparrow;-\varphi\rangle'\Bigr)+\sin\Delta\Bigl(|k^a_+,\uparrow;\varphi\rangle'|k^a_-,\uparrow;-\varphi\rangle'+|k^a_+,\downarrow;\varphi\rangle'|k^a_-,\downarrow;-\varphi\rangle'\Bigr)\Bigr] \, ,
\ee
where
\be\label{delta}
\Delta=\Theta-\varphi = \varphi\,\Bigl[\cosh\zeta\lf(1+\frac{rg'}{2g}-\frac{rf'}{2f}\ri)-1\Bigr] \, .
\ee
Before we can evaluate the CHSH inequality~\eqref{CHSHQuant} for the observables introduced in Eq.~\eqref{Observables}, the measurement operators must be rewritten in the new reference frame obtained as a result of the rotation, that is
\begin{equation}
\label{ObservablesPrime}
\begin{aligned}
\hat{A}_1^\prime =\cos{\Theta }\hat{\Si}_x^{(A)} - \sin{\Theta} \hat{\Si}_z^{(A)}&\,, \quad \hat{A}_2^\prime =\hat{\Si}_y^{(A)}\,, \\[2mm]
\hat{B}_1^\prime = -\frac{\cos{\Theta }\left( \hat{\Si}_x^{(B)} + \hat{\Si}_y^{(B)} \right) + \sin{\Theta} \hat{\Si}_z^{(B)} }{\sqrt{2}}&\,, \quad \hat{B}_2^\prime =\frac{ \cos{\Theta } \left( \hat{\Si}_x^{(B)} - \hat{\Si}_y^{(B)} \right)+ \sin{\Theta} \hat{\Si}_z^{(B)} }{\sqrt{2}} \, .
\end{aligned}
\end{equation}
{Note that, in terms of experimental complexity, the preparation of the detectors does not require difficult steps. Indeed, as the observers lie in a locally inertial reference frame, the only required additional action with respect to a standard CHSH test would consist in rotating the experimental apparatus so as to match the angular distance spanned by the pair of particles (that is, $\pm\varphi$). However, this distance is known \emph{a priori} once the endpoints (and thus the location of the observers) have been determined on the circular orbit.}

Collecting all the above results, we finally obtain
\begin{equation}\label{coreq}
\mathcal{S}^\prime[\vert \psi^\prime \rangle] = 2 \sqrt{2} \cos^2 \Delta \, .
\end{equation}
We can interpret Eq.~\eqref{coreq} as follows: when a CHSH-like experiment is carried out after the observables have been rotated, the maximal initial violation of the CHSH inequality $2\sqrt{2}$ becomes modulated by a factor $\cos^2 \Delta$. 

Inspecting Eq.~\eqref{delta}, we see that, in the presence of the gravitational interaction, the phase shift parameter $\Delta$ responsible for the overall violation of the CHSH inequality acquires contributions that depend on the details of the spacetime in which the entangled particles propagate. Therefore, depending on the actual nature of the ultra-compact gravitational source being considered, we expect to find distinct and possibly significantly different modulations in the violation of the CHSH inequality. {A comment is in order here: since the action of the Wigner rotation on the initial state is but a local unitary map, the total degree of nonlocality should not be influenced by it. As a matter of fact, one can check that, with a suitable selection of different directions for the observables \cite{ueda}, the perfect EPR correlation would be restored even after the propagation of the two spins along the orbit. Hence, the effect of gravity and of the external acceleration required to maintain a circular trajectory only amounts to changing the orientation of the measurements; in other words, nonlocality remains essentially unaffected.}

{On a final note, it is worth observing that, as long as the initial state is prepared in close analogy to the one introduced in Eq. \eqref{instate} and the ensuing observables are chosen in such a way to reproduce the maximally allowed degree of nonlocality, no qualitative deviation from the current analysis is expected to arise when a different pair of particles with arbitrary spins is used to study the problem at hand.}


\section{Comparing models of ultra-compact objects with Bell nonlocality}

The gravitational modulation of the Bell nonlocality derived in the previous Section, i.e., 
Eqs.~\eqref{delta} and \eqref{coreq}, can be exploited to compare relevant alternative models of ultra-compact structures, such as the string fuzzball and the black hole (classical or with perturbative quantum corrections). {We would like to stress again that these equations are completely general, and can thus be applied to an arbitrary black hole-like solution whose metric can be cast in the form \eqref{metric}.} 

\subsection{String fuzzballs}

Within string-theory inspired cosmology, fuzzballs~\cite{fuzz1,fuzz2,rew} are spheres of strings of definite, finite volume that simulate the behavior of black holes, but having the two main problems plaguing the latter (the singularity and the information paradox) removed by the finite length extension of their microscopic components. 

A concrete fuzzball solution amenable to quantitative investigation is obtained from $\mathcal{N}=2$ four-dimensional supergravity, with a non-minimal coupling between gravity, four $U(1)$ gauge fields and three complex scalars. This particular case allows for some explicit phenomenological predictions that might soon be tested via gravitational waves detection by studying ringdown (gravitational wave peak in merging events), quasi-normal modes, and spectroscopy~\cite{bianchi}.

In isotropic spherical coordinates, a four-dimensional fuzzball geometry can be described by the line element~\cite{bianchi}
\be\label{fuzz}
ds^2=-\frac{dt^2}{\sqrt{H_1H_2H_3H_4}}+\sqrt{H_1H_2H_3H_4}\lf[dr^2+r^2d\theta^2+r^2\sin^2\theta d\varphi^2\ri] \, ,
\ee
where $H_A=1+Q_A/r$, $A=1,2,3,4$, with $Q_A$ being electric and magnetic charges. The total mass of the fuzzball is given by $M=\lf(Q_1+Q_2+Q_3+Q_4\ri)/4$, and one recovers the extremal Reissner-Nordstr\"om black hole solution when all the charges are equal. 
A straightforward comparison with Eq.~\eqref{metric} yields the identification 
\be\label{fuzz2}
f(r)=\frac{1}{g(r)}=\frac{1}{\sqrt{H_1H_2H_3H_4}} \, .
\ee
{Before concluding, we observe that, for the sake of comparing different spherically-symmetric scenarios, we consider a metric that does not include ``tidal'' effects due to the non-overlapping of the three centers with which fuzzball multicenter microstate solutions are built in \cite{bianchi}. Effectively, this amounts to considering only the metric of the extremal black hole solution. However, this approximation becomes more precise the more massive the compact object is, since the impact of tidal forces on nearby test masses scales as the inverse of the mass \cite{spag}. Therefore, by focusing on supermassive gravitational sources, we could actually regard such effects as negligible from a physically sound perspective.}

\subsection{Classical and quantum Schwarzschild black holes}

Black holes can be investigated both in a general-relativistic classical context as well as in a quantum-corrected one. The last instance occurs when one considers gravity as an effective field theory, so that quantum gravitational radiative corrections influence the energy-momentum tensor appearing in Einstein equations. In turn, such a modification gives rise to long-range corrections appearing in the expression of the metric tensor $g_{\mu\nu}$.
In general, the magnitude of such corrections is extremely small and can be neglected, but in the proximity of a black hole they actively affect the metric and the ensuing gravitational phenomenology. Therefore, we can investigate the implications of the CHSH experiment in the strong-gravity regime, both in the classical and in the quantum-corrected framework. 

\begin{figure}[ht]
  \centering
    \includegraphics[width=15.5cm]{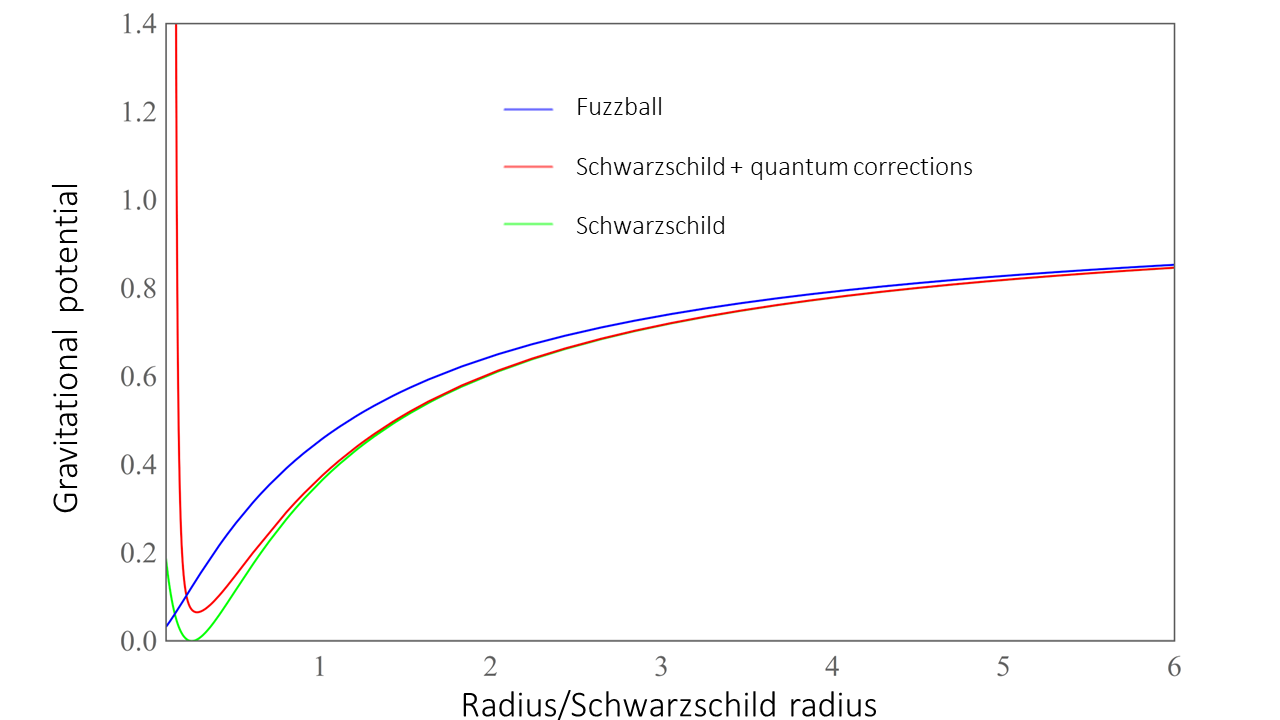}
    \caption{The gravitational potential of an ultra-compact object as a function of the distance (radius) from the object center in units of the Schwarzschild radius for different spacetimes. Sample values are fixed at $M=2.5$, $Q_1=1$, $Q_2=2$, $Q_3=3$, and $Q_4=4$.}
  \label{figure2}
\end{figure}

Specifically, we are interested in the extrapolation of the Schwarzschild-like solution inclusive of quantum corrections in the isotropic coordinate system~\cite{don}. The line element associated with the quantum-corrected spacetime reads
\be\label{qsc}
ds^2=-\lf[\frac{1-\frac{M}{2r}+\frac{31}{30\pi}\frac{M}{r^3}}{1+\frac{M}{2r}-\frac{31}{30\pi}\frac{M}{r^3}}\ri]^2dt^2+\lf(1+\frac{M}{2r}-\frac{7}{30\pi}\frac{M}{r^3}\ri)^4\lf[dr^2+r^2d\theta^2+r^2\sin^2\theta d\varphi^2\ri]\,, 
\ee
where $M$ is the mass of the black hole. The standard Schwarzschild solution is recovered when the additive corrections that depend on $1/r^3$ in Eq. \eqref{qsc} are removed. By comparing the metric of Eq.~\eqref{qsc} with the one of Eq.~\eqref{metric}, the following identification holds:
\be\label{qsc2}
f(r)=\lf[\frac{1-\frac{M}{2r}+\frac{31}{30\pi}\frac{M}{r^3}}{1+\frac{M}{2r}-\frac{31}{30\pi}\frac{M}{r^3}}\ri]^2\,, \qquad
g(r)=\lf(1+\frac{M}{2r}-\frac{7}{30\pi}\frac{M}{r^3}\ri)^4\,.
\ee

\subsection{Comparison}

We can now compare the different ultra-compact objects and establish if and how the fuzzball and black hole solutions differ in their response to the CHSH quantum nonlocality test. 

\begin{figure}[ht]
  \centering
    \includegraphics[width=15.5cm]{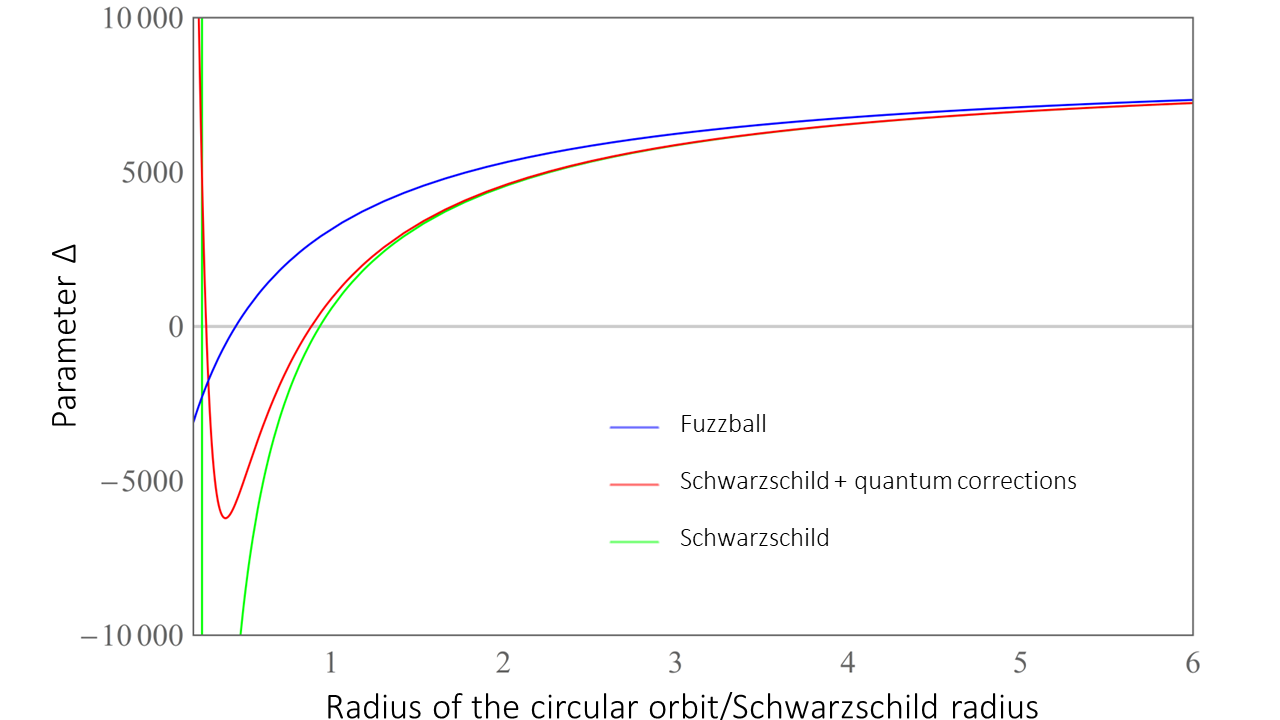}
    \caption{The magnitude of the gravitational modulation parameter $\Delta$ as a function of the radius of the circular orbit in units of the Schwarzschild radius for different spacetimes. Blue solid line: $\Delta_{SF}$, Eq.~\eqref{deltafuzz}, corresponding to the string fuzzball solution. Red solid line: $\Delta_{QS}$, Eq.~\eqref{deltaqsc}, corresponding to the quantum-corrected Schwarzschild solution. Green solid line: $\Delta_{SF}$, Eq.~\eqref{deltacs}, corresponding to the classical Schwarzschild solution.  Sample values are fixed at $\varphi=\pi/4$, $M=2.5$, $Q_1=1$, $Q_2=2$, $Q_3=3$,  $Q_4=4$ and $\zeta=10$.}
  \label{figure3}
\end{figure}

Firstly, we observe that the two classes of objects already differ at the classical level in the behavior of the gravitational potential in regions sufficiently close to the event horizon, as illustrated in Fig.~\ref{figure2}.
Next, in order to estimate quantitatively the distinct predictions in the quantum regime, we need to evaluate the gravitational modulation parameter of Bell nonlocality $\Delta$, i.e., Eq.~\eqref{delta}, in each case. Since we have specialized the general line element~\eqref{metric} to the instances~\eqref{fuzz} and~\eqref{qsc}, we are left with the task of taking advantage of the expressions of $f(r)$ and $g(r)$ appearing in Eqs.~\eqref{fuzz2} and~\eqref{qsc2} and derive the explicit form of $\Delta$. 

\begin{figure}[ht]
  \centering
    \includegraphics[width=15.5cm]{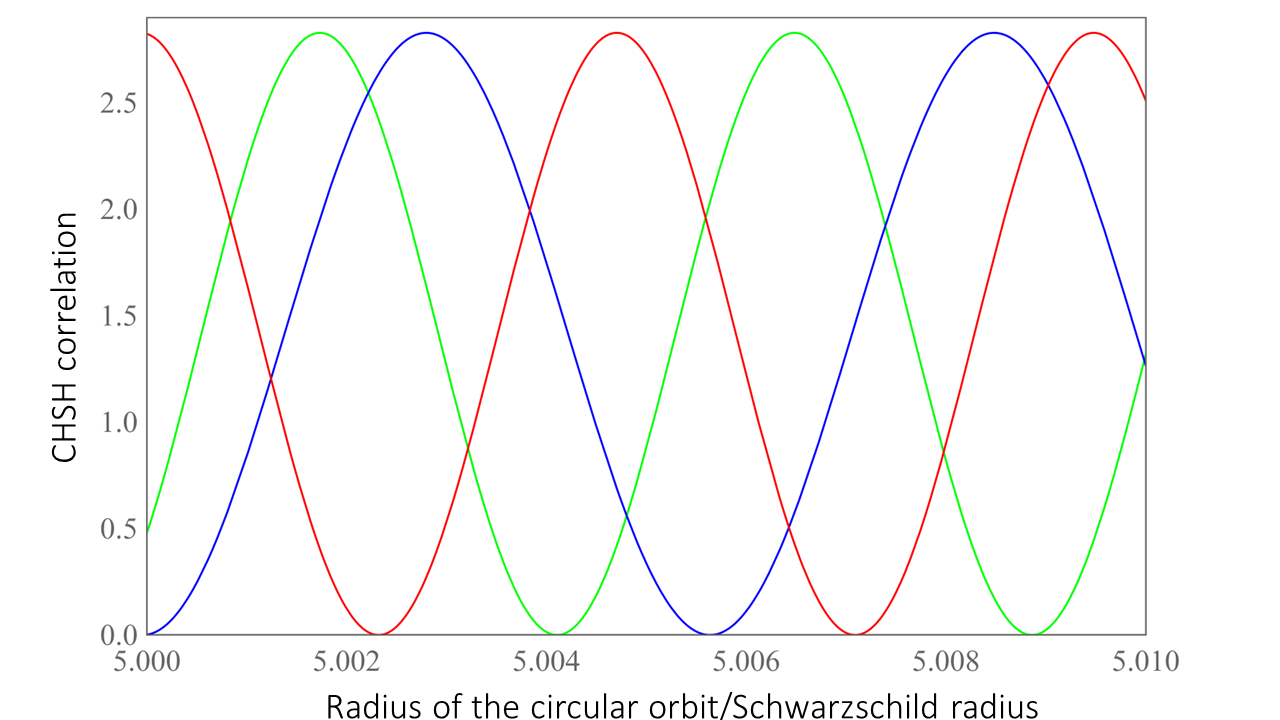}
    \caption{CHSH correlation as a function of the orbit radius in units of the Schwarzschild radius for different spacetimes. Blue solid line: string fuzzball solution. Red solid line: quantum-corrected Schwarzschild solution. Green solid line: classical Schwarzschild solution. Sample values are fixed at $\varphi=\pi/4$, $M=2.5$, $Q_1=1$, $Q_2=2$, $Q_3=3$,  $Q_4=4$ and $\zeta=10$.}
  \label{figure4}
\end{figure}

As a preliminary step, we determine the form $\Delta_{CS}$ of the parameter $\Delta$ holding for the isotropic, classical standard Schwarzschild solution, that is
\be\label{deltacs}
\Delta_{CS}=\varphi\lf\{\frac{\cosh\zeta}{2}\frac{M^2+10Mr-8r^2}{\lf(M-2r\ri)\lf(M+2r\ri)}-1\ri\}\,.
\ee
Accounting for the quantum perturbative corrections appearing in Eq.~\eqref{qsc}, one can derive the form $\Delta_{QS}$ of the parameter $\Delta$ holding for the quantum-corrected Schwarzschild solution at leading order
\be\label{deltaqsc}
\Delta_{QS}=\varphi\lf\{\frac{\cosh\zeta}{2}\frac{\beta(r)}{\lf[30\pi r^3+M\lf(31-15\pi r^2\ri)\ri]\lf[30\pi r^3-M\lf(31-15\pi r^2\ri)\ri]\lf[30\pi r^3-M\lf(7-15\pi r^2\ri)\ri]}-1\ri\}\,,
\ee
where
\be\label{beta}
\beta(r)=54000\pi^3r^9-M^3\lf(31-15\pi r^2\ri)^2\lf(79+15\pi r^2\ri)+900M\pi^2r^6\lf(451-45\pi r^2\ri)-60M^2\pi r^3\lf(2263-3930\pi r^2+675\pi^2r^4\ri) \, .
\ee
Finally, making use of the potentials $f(r)$ and $g(r)$ in Eq.~\eqref{fuzz2}, we obtain the form $\Delta_{SF}$ of the parameter $\Delta$ holding for the string fuzzball solution
\be\label{deltafuzz}
\Delta_{SF} = \varphi\lf\{\frac{\cosh\zeta}{2}\frac{r^3\lf(Q_3+Q_4+2r\ri)+Q_2\lf(r^2-Q_3Q_4\ri)r-Q_1\lf[Q_3Q_4r-r^3+Q_2\lf(2Q_3Q_4+Q_3r+Q_4r\ri)\ri]}{(Q_1+r)(Q_2+r)(Q_3+r)(Q_4+r)}-1\ri\}\,.
\ee
The three expressions $\Delta_{CS}$, $\Delta_{QS}$ and $\Delta_{SF}$ {and the ensuing modulations} differ significantly when evaluated along orbits sufficiently close to the ultra-compact objects, as illustrated in Fig.~\ref{figure3} {and Fig.~\ref{figure4}, respectively}. We see that, in the strong-gravity regime, the specific nature of the gravitational source affects dramatically the degree of violation of the CHSH inequality; hence, the gravitational modulation of Bell nonlocality allows to distinguish different models of ultra-compact objects and to discriminate between the string theory and quantum field theory approaches to quantum gravity.


{In particular,} in Fig.~\ref{figure4} we plot the oscillatory modulation of Bell nonlocality as measured by the degree of violation of the CHSH inequality for orbits close to the event horizon, that is, for an interval of the radial coordinate that does not exceedingly deviate from the Schwarzschild radius. For a wider range of values of the radial coordinate, the frequency of the oscillations $\cos^2\Delta$ grows very rapidly, thereby blurring the interpolation patterns. Moreover, as the quantities $\Delta_S$, $\Delta_{QS}$ and $\Delta_F$ share the same behavior in the limit $r\gg2M$, the phase shifts of the CHSH correlations can no longer be resolved in this regime. 

{In connection with the above reasoning, it is important to recall that, by selecting a given circular trajectory, we are essentially fixing the value of the radius. Therefore, the figures and the modulating parameters do not really have to be interpreted as functions of $r$, because once its value is chosen they cannot vary with the particles' dynamical evolution. Hence, the evaluation of the differences between the predictions of the distinct models has to be intended for a fixed value of the radius. For instance, by looking at Fig.~\ref{figure4}, it is immediate to verify that, for certain values of $r$ (\emph{i.e.}, for certain circular orbits), the magnitude of $\mathcal{S}'$ is exactly zero according to some models, thereby signaling a complete absence of correlations between the two spins. For these exact values, instead, the other models predict a non-vanishing degree of CHSH correlations, thus entailing that the outcome of the gedanken experiment is unambiguously distinguishable.}


\section{Discussion}

We have discussed a gedanken experiment that shows how quantum nonlocality in strong gravitational fields leads to predictions that discriminate between different models of quantum gravity phenomenology, both perturbative and non-perturbative ones. CHSH nonlocality tests with entangled particle pairs on circular orbits near ultra-compact objects show that the spin precession occurring in curved spacetime is responsible for a modulation of the degree of quantum nonlocality. 
In the presence of a non-trivial spacetime background, the standard maximally allowed violation $2\sqrt{2}$ of the CHSH inequality becomes $2\sqrt{2}\cos^2\Delta$, with the angular modulation factor $\Delta$ strictly dependent on the metric tensor components and heavily influenced by the conjectured underlying nature of the ultra-compact object considered. A simple measurement of quantum nonlocality can thus be used to validate or falsify some phenomenological models of quantum gravitational effects in the strong-field regime. 

The thought experiment we have conceived provides further evidence supporting the use of quantum information concepts like entanglement and Bell nonlocality as a key tool in the investigation of yet hypothetical quantum gravitational phenomena. In deriving the main result of our work, we have also generalized the formalism of Refs.~\cite{ueda,our} on the study of quantum correlations in gravitational fields, so as to make it applicable in general, beyond the Schwarzschild solution and weak-field limit in the isotropic coordinate system. 

Concerning possible experimental tests probing the quantum nature of the gravitational field, while high-energy scattering processes are still too far from the experimental scales needed to detect coexisting quantum and gravitational effects, it is a largely shared belief that tiny signatures of such a coexistence might still be revealed with currently available means through satellite experiments, gravitational wave analysis or tabletop laboratory tests centered around foundational aspects of quantum mechanics. In this respect, it is worth remarking that credited proposals which aim at collecting a direct measurement of quantum gravity phenomenology are essentially based either upon decoherence/gravity-based wave function collapse models~\cite{d1,d2,d3,d4,d5,d6,d7,d8,d9,d10} or, as already mentioned, the detection of gravitationally-induced quantum correlations by quantum gravitational mediators~\cite{Bose,Marletto,Schmole,Carney,Pedernales,Weiss,Cosco,Christodoulou,Bosso}.
The hard challenge facing the experimental implementation of these table-top laboratory tests is that of realizing correlated and delocalized superpositions of large enough masses. So far, spatial superpositions have been observed with masses at most of the order of $10^{-23}$ Kg (large molecules) \cite{Fein}, while the faintest gravitational field that can be currently measured is the one generated by masses of the order of $10^{-4}$ Kg \cite{Westphal}. 

While one may hope to significantly improve these numbers by considering some amplification mechanisms, table-top probing of gravitational effects on Bell inequality might turn out to be significantly less challenging and might open the way to laboratory simulations of extreme cosmological conditions suitable for the verification of quantum-gravity-induced modulations of quantum nonlocality as described in the present work. In this respect, a particularly promising avenue might involve designing experimental tests of the CHSH inequality near the horizon of sonic and optical analogues of black holes \cite{Visser,Unruh,Philbin,Kolobov}.

On a final note, our findings strongly suggest that resorting to the entire spectrum of quantum resources, from nonlocality, entanglement and steering to discord, coherence and complementarity may provide very useful insights in the investigation on the actual nature of gravity. 

\section*{Acknowledgements}

The authors acknowledge support by MUR (Ministero dell’Universit\`a e della Ricerca) via the project PRIN 2017 ``Taming complexity via QUantum Strategies: a Hybrid Integrated Photonic approach'' (QUSHIP) Id. 2017SRNBRK. L. P. acknowledges networking support by the COST Action CA18108 and financial support from the ``Angelo Della Riccia'' Foundation. {The authors would like to thank the anonymous reviewers for their suggestions and comments which helped to improve the quality of the manuscript.}

\end{document}